\documentclass[12pt,preprint]{aastex}

\begin{document}

\title{A Dark Hydrogen Cloud in the Virgo Cluster}
\author{
Robert Minchin\altaffilmark{1}, 
Jonathan Davies\altaffilmark{1}, 
Michael Disney\altaffilmark{1}, 
Peter Boyce\altaffilmark{2}, 
Diego Garcia\altaffilmark{1}, 
Christine Jordan\altaffilmark{3}, 
Virginia Kilborn\altaffilmark{4}, 
Robert Lang\altaffilmark{1}, 
Sarah Roberts\altaffilmark{1}, 
Sabina Sabatini\altaffilmark{5} 
\and 
Wim van Driel\altaffilmark{6}}
\affil{}

\altaffiltext{1}{School of Physics and Astronomy, Cardiff University, Cardiff,
CF24 3YB, UK; Robert.Minchin@astro.cf.ac.uk, Jonathan.Davies@astro.cf.ac.uk,
Mike.Disney@astro.cf.ac.uk, Diego.Garcia@astro.cf.ac.uk, LangRH@cardifff.ac.uk,
Sarah.Roberts@astro.cf.ac.uk}
\altaffiltext{2}{Planning Division, Cardiff University, Park Place, Cardiff, 
CF10 3UA, UK; BoyceP@cardiff.ac.uk}
\altaffiltext{3}{Jodrell Bank Observatory, University of Manchester, 
Macclesfield, Cheshire, SK11 9DL, UK; caj@jb.man.ac.uk}
\altaffiltext{4}{Centre for Astrophysics and Supercomputing, Swinburne 
University of Technology, Mail 31, P.O. Box 218, Hawthorn, Victoria 3122, 
Australia; vkilborn@astro.swin.edu.au}
\altaffiltext{5}{Osservatorio Astronomico di Roma, via Frascati 33, I-00040,
Monte Porzio, Italy; sabatini@mporzio.astro.it}
\altaffiltext{6}{Observatoire de Paris, GEPI, CNRS UMR 8111 and Universit\'e 
Paris 7, 5 place Jules Janssen, F-92195 Meudon Cedex, France; 
Wim.vanDriel@obspm.fr}

\shorttitle{A Dark H\,{\sc i} Cloud in the Virgo Cluster}
\shortauthors{Minchin et al.}
 
\begin{abstract}
VIRGOHI21 is an H\,{\sc i} source detected in the Virgo Cluster survey of
\citet{Davies04} which has a neutral hydrogen mass of $10^8 M_\odot$ and
a velocity width of $\Delta V_{20} = 220$ km\,s$^{-1}$.  From the Tully-Fisher
relation, a galaxy with this velocity width would be expected to be 12th
magnitude or brighter; however deep CCD imaging has failed to turn up a 
counterpart down to a surface-brightness level of 27.5 $B$\,mag\,arcsec$^{-2}$.
The H\,{\sc i} observations show that it is extended over at least 16 kpc 
which, if the system is bound, gives it a minimum dynamical mass of 
$\sim 10^{11} M_\odot$ and a mass to light ratio of $M_{dyn}/L_{B}>500 
M_\odot/L_\odot$.  If it is tidal debris then the putative parents have 
vanished; the remaining viable explanation is that VIRGOHI21 is a dark halo 
that does not contain the expected bright galaxy.  This object
was found because of the low column density limit of our survey, a limit much
lower than that achieved by all-sky surveys such as HIPASS.  Further such
sensitive surveys might turn up a significant number of the dark matter halos 
predicted by Dark Matter models.
\end{abstract}

\keywords{dark matter --- galaxies: general --- galaxies: clusters: individual 
(Virgo) --- radio lines: galaxies}

\section{Introduction} 

Simulations of Cold Dark Matter (CDM) models predict far more dark matter
halos than are observed as galaxies \citep{Klypin99,Moore99}.  For this
reason, it has been hypothesized that there must exist dark matter halos that 
contain no stars \citep[e.g.][]{Jimenez97,Verde02}.  The advent of neutral
hydrogen multibeam systems has allowed surveys of large areas of sky to be 
carried out with much higher sensitivity than has been possible in the past, 
thus allowing sources to be detected by their gas content alone rather than 
their
stars and opening up the possibility of finding truly isolated clouds of 
extragalactic gas with no stars.  Prior to this, blind H\,{\sc i} surveys
either covered very small areas or were
insensitive to H\,{\sc i} column densities lower than $\sim 10^{20}$ cm$^{-2}$
($\sim 1 M_\odot$ pc$^{-2}$) \citep{Minchin03}.

\citet{Davies04} used the multibeam system on the Lovell telescope at Jodrell
Bank Observatory to carry  out a deep neutral-hydrogen (H\,{\sc i}) survey of 
the Virgo Cluster (VIRGOHI), covering 32 square degrees and detecting 31
sources. Of these sources, 27 were known cluster members and 4 were new 
detections.  One of the latter lay behind M86 and was thus unobservable
optically and one was undetected in follow-up observations and is therefore 
believed to 
be a false detection.  The other two were confirmed at Arecibo and flagged by 
Davies et al. as possible isolated H\,{\sc i} clouds.  One (VIRGOHI27) has
an optical counterpart visible in our deep CCD images, the other (VIRGOHI21,
the subject of this letter) does not.

There have been several previous claims of the detection of isolated clouds of 
extragalactic gas with no stars in them, but subsequent analyses have either 
revealed the optical counterparts \citep{Giovanelli89,McMahon90}, or 
shown that the gas is merely debris from 
nearby visible galaxies \citep{Schneider83}.  Many other detections of 
H\,{\sc i} clouds have been associated with nearby optically-bright 
galaxies \citep{Kilborn00,Boyce01,Ryder01}.  VIRGOHI21 cannot be so easily 
explained.

\section{Further observations}

Following detection, VIRGOHI21 was re-observed at Arecibo.  The observations 
are
fully described by \citet{Davies04}, here we give a much more detailed analysis
of the Arecibo data and present new VLA and optical observations.  The Arecibo 
observations used a number of pointings in a
pattern around the best-fit location from the Jodrell Bank data, leading to
the source being detected in five of the Arecibo beams.

\begin{figure}
\plotone{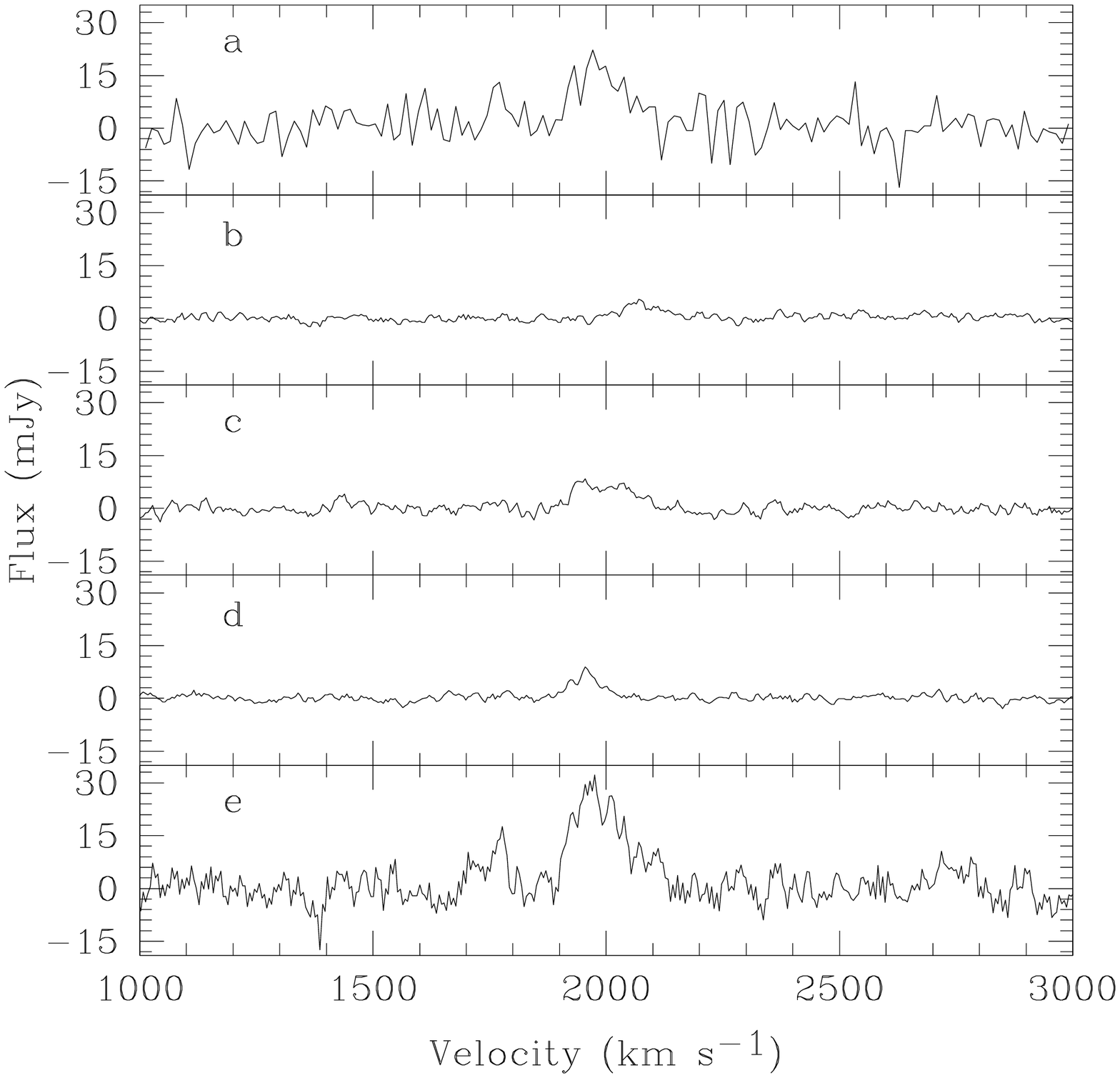}
\caption{Atomic Hydrogen spectra of VIRGOHI21 plotted as a function of
recessional velocity.  (a) discovery spectrum from Jodrell Bank (3500 s).
(b) - (d) Arecibo follow-up spectra from beams labelled (b) - (d) in
Fig. \ref{pointing}, cutting north-south across the source (600 s). (e) Sum
of all Arecibo spectra (600 s per beam).  The peak at $\sim
1750$ km\,s$^{-1}$ in spectrum (e) can be clearly associated with galaxy
C on Fig. \ref{int} by our VLA data.}
\label{spectra}
\end{figure}

\begin{figure}
\plotone{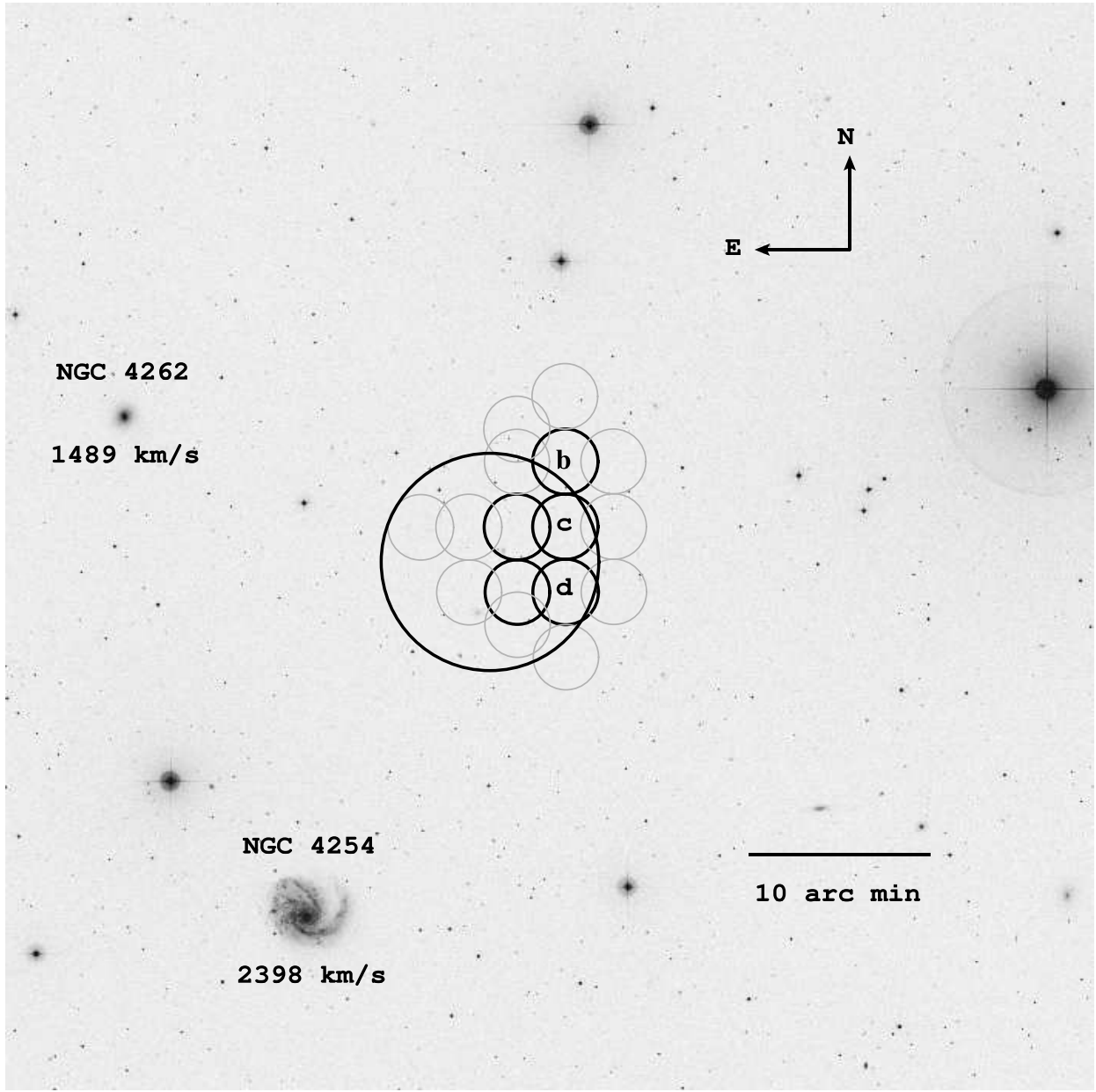}
\caption{The Arecibo pointing pattern for VIRGOHI21 overlaid on a digitized 
sky survey image of the region around the source.  Circles mark the beam 
position and FWHM; black circles denote beams where a firm detection (better 
than $4\sigma$) was made
and grey circles those beams which do not contain a definite signal.  The 
large circle marks the beam corresponding to the pixel in the Jodrell Bank data
cube from which the spectrum and 
measurements in Fig. \ref{spectra} (a) are taken.  The beams marked (b), (c) 
and (d) correspond to the spectra marked (b), (c) and (d) in Fig. 
\ref{spectra}}
\label{pointing}
\end{figure}

Fig. \ref{spectra} shows the single-dish spectra of VIRGOHI21.  Spectrum (a) is
the discovery spectrum from Jodrell Bank, this has a noise level of 4 mJy
per 13.2 km\,s$^{-1}$ channel and a $5\sigma$ column-density sensitivity
$N_{HI,lim} = 7\times 10^{18}$ cm$^{-2}$ if spread over 200 km\,s$^{-1}$.  From
it we 
measure a total flux of $F_{HI} = 2.4 \pm 0.3$ Jy km\,s$^{-1}$ and a velocity 
width at 20 per cent of the peak flux of $\Delta V_{20} = 290$ km\,s$^{-1}$.  
Spectra (b), (c) and (d) are three north through south beams across the source 
from the Arecibo observations (labelled (b) - (d) in Fig. \ref{pointing}).  
These have a noise level of 1.3 mJy per 5.5 km\,s$^{-1}$ channel, giving 
$N_{HI,lim} = 2.7\times 10^{19}$ cm$^{-2}$ over 200 km\,s$^{-1}$.  They 
reveal a systematic increase in velocity of $\sim 200$ 
km\,s$^{-1}$ from south to north.  Spectrum (e) is the co-added spectrum from 
all 16 Arecibo beams (shown in Fig. \ref{pointing}) from which we measure 
$F_{HI} = 3.8 \pm 0.2$ Jy km\,s$^{-1}$ and $\Delta V_{20} =  220$ km\,s$^{-1}$

Fig. \ref{pointing} shows the Arecibo pointing pattern for VIRGOHI21 and which 
beams made firm detections (better than $4\sigma$).  It can be seen that 
VIRGOHI21 is extended over at least one
Arecibo beam width $\approx 3.6$ arc min, or 16 kpc at an assumed distance to 
Virgo of 16 Mpc \citep{Graham99}.  Using the measured H\,{\sc i} flux, we 
calculate an H\,{\sc i} mass of $2\times 10^8 M_\odot$ if it is
at the distance of the Virgo Cluster, or $7\times 10^8 M_\odot$ if it is at
its Hubble distance (29 Mpc for H$_0$ = 70 Mpc$^{-1}$ km\,s$^{-1}$).  For the 
rest of this letter we will assume the former as more conservative in this
context.  The best-position for the 
center of the H\,{\sc i} emission, formed by weighting the Arecibo detection 
positions by their fluxes, is $12^h17^m53.6^s, 
+14^\circ45^\prime25^{\prime\prime}$ (J2000).  We can dismiss the possibility 
that this is side-lobe emission from another part of the sky, because VIRGOHI21
has been detected with two telescopes with very different side-lobes.  
Additionally, there are no H\,{\sc i}-massive galaxies in the region that match
its velocity profile \citep{Davies04}.  

H\,{\sc i} observations with the VLA in D-array in August 2004 reached a
$5\sigma$ column-density limit of $10^{20}$ cm$^{-2}$ over 60 km\,s$^{-1}$ in
6 hours (0.5 mJy per 20 km\,s$^{-1}$ channel with a beam size of $48 \times 45$
arc seconds using natural weighting).  Solar interference on the shorter 
baselines meant that the 
observations did not reach the hoped-for column-density sensitivity; most
of the data from baselines shorter than $\sim 270$\,m ($\sim 1.3 $k$\lambda$)
had to be flagged as bad.  These observations did detect compact 
H\,{\sc i} associated with a nearby dwarf elliptical (MAPS-NGP O\_435\_1291894)
at a different velocity, but, although sensitive to compact, narrow-line gas
down to a $5\sigma$ limit of $3\times 10^6 M_\odot$, were not sensitive 
enough to low column-density, high velocity-width
gas to detect VIRGOHI21.  For the Arecibo and VLA observations to be 
consistent, the source must again exceed 16 kpc in diameter.

We have obtained deep optical CCD images in $B$, $r$ and $i$
bands with the 2.5-m Isaac Newton Telescope (INT).  By binning the $B$-band
image (generally the best band for looking for low surface brightness galaxies)
into 1 arc sec pixels, we reach a surface-brightness limit of 27.5 $B$ 
mag arcsec$^{-2}$; treating the $r$ and $i$ images in the same way gives
surface-brightness limits of $\sim 27.0$ and of 25.8 mag arcsec$^{-2}$
respectively.  Previous experience indicates that, on the $B$-band frame,
we should be able to easily detect objects of ten arc sec scale or larger at 
this surface brightness limit \citep{Sabatini03,Roberts04}.  This is more 
than 100 times dimmer than the central surface brightness of the disks of 
typical spiral galaxies \citep[21.5 $B$ mag arcsec$^{-2}$,][]{Freeman70} and 
dimmer than any known massive low surface-brightness galaxy \citep[26.5 $B$
mag arcsec$^{-2}$,][]{Bothun87} or (for typical $B - V$ colors of $\sim 0.6$) 
than the lowest surface-brightness dwarf galaxy \citep[26.8 $V$ mag 
arcsec$^{-2}$][]{Zucker04}.

\begin{figure}
\plotone{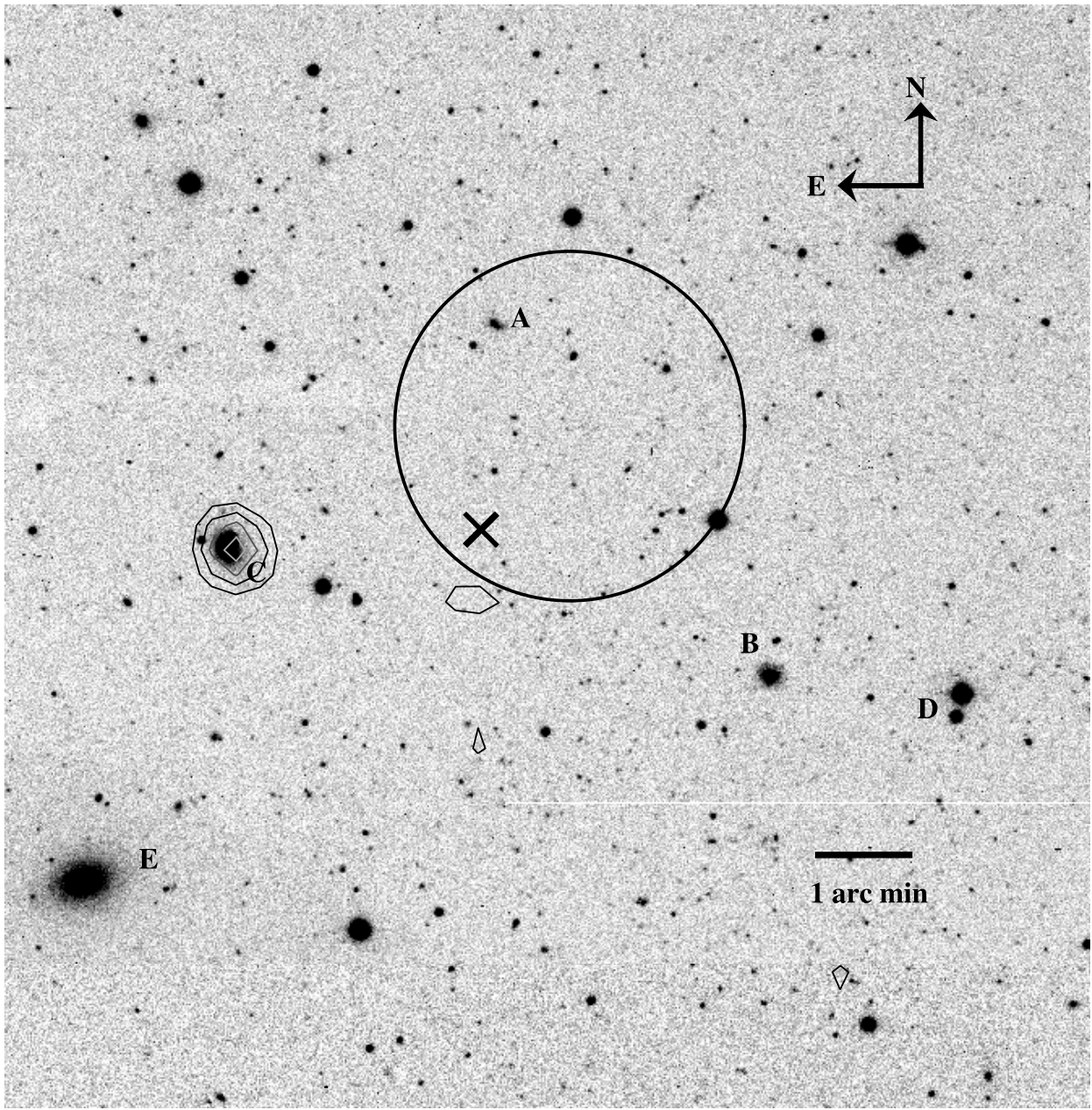}
\caption{Isaac Newton 
Telescope $B$-band optical image (shown as negative) of the field of 
VIRGOHI21.  The cross marks the weighted center of the H\,{\sc i} detection
and the circle shows the size and position of the central Arecibo beam.
Optical sources labelled (A) -- (E) are discussed in the text.  Contours show
the moment map from the VLA detection of source (C) ($M_{HI} = 1.4\times 10^7 
M_\odot$) at 1750 km\,s$^{-1}$ and are at 3, 6, 9 and 12 sigma.}

\label{int}
\end{figure}

Although we easily were able to identify an optical counterpart (at $12 ^h26 
^m40.1^s$, $+19^\circ 45^\prime 50^{\prime\prime}$ J2000) to the other 
possible H\,{\sc i} cloud, VIRGOHI27, no optical counterpart to VIRGOHI21 is 
visible down to our surface-brightness limit on any of our deep images of this 
region (Fig. \ref{int}), nor can one be found with advanced routines for 
detecting low surface-brightness galaxies \citep[matched filtering and 
wavelets,][]{Sabatini03}.  Unlike VIRGOHI27, the bluest 
objects in the field, which might be associated with H\,{\sc ii} regions, are 
widely distributed without any concentration towards the H\,{\sc i} center.
Looking at the statistics of the sky noise for the frame, the mean number of 
counts and the standard deviation (excluding stars) are similar in the area of 
the H\,{\sc i} detection to other, blank areas of sky in the vicinity.  The 
number of detected faint objects is not significantly above the
average in a box centred on the H\,{\sc i} position: there are 3 objects with
$m_B > 23$ in a $100\times 100$ pixel ($33.3\times 33.3$ arc sec) region
centred on our best H\,{\sc i} position, compared to an average of $1.8 \pm 
1.4$ across the cube, and 4 objects with $m_B > 22$ compared to an average of
$2.6 \pm 1.7$.

As is to be expected there are some features on the image of VIRGOHI21 that are
obviously faint galaxies. These are labelled (A) - (E) in Fig. \ref{int}.  
(A) is a small source with a star superposed (which prevents accurate 
determination of its color and luminosity) 2 arc minutes north of the 
weighted center, just within the Full Width Half Maximum (FWHM) of the 
strongest H\,{\sc i} beam.  Its east-west orientation is at odds with the 
north-south orientation expected from the velocity field of VIRGOHI21 and, if
VIRGOHI21 is rotating, this galaxy is too far north for the rotation to be 
centred on its position.  Neither its size, position nor orientation make
this a likely optical counterpart to VIRGOHI21.  (B) is 
another uncatalogued galaxy, 3.5 arc minutes south-west of the weighted center.
It lies within the FWHM of an Arecibo beam where there was no detection, thus 
it cannot be the optical counterpart.  There are three objects classified as 
galaxies within six arc minutes.  One (C) at the very edge of one of the 
Arecibo beams is a dwarf elliptical (MAPS-NGP O\_435\_1291894) that is detected
in our VLA data at 1750 km\,s$^{-1}$ (see Fig. \ref{spectra}).  The H\,{\sc i}
is separated both spatially and in velocity from VIRGOHI21 and so it cannot 
therefore be the optical counterpart.
Another of the catalogued galaxies (D - MAPS-NGP O\_435\_1292289)
is a double star miscatalogued as a galaxy, and the third (E - VCC 0273) lies 
in an Arecibo beam where no detection was made.

We conclude that there is no optical counterpart to VIRGOHI21 
down to a $B$-band surface-brightness limit of 27.5 $B$ mag arcsec$^{-2}$.  
This is less than 1 solar luminosity pc$^{-2}$, giving a maximum luminosity in 
stars of $\sim 10^8$ solar luminosities if a diameter of 16 kpc is assumed. If 
VIRGOHI21 is a bound system (see discussion below), this leads to a mass to 
light ratio in solar units of 
$M_{dyn}/L_{B}>500$ compared to a typical $L^\star$ galaxy like the Milky
Way with $M_{dyn}/L_{B} \sim 50$ within its H{\sc i} radius \citep{Salucci97}. 
For standard stellar $M/L_B$ ratios the upper limit on the mass in stars is 
approximately equal to the mass in H\,{\sc i}.

\section{Discussion}

The closest bright H\,{\sc i}-rich galaxies ($M_B < -16$, $M_{HI} > 10^8 
M_\odot$) are shown in Fig. \ref{pointing}.  These are NGC 4262 at 
1489 km\,s$^{-1}$ and NGC 4254 at 2398 km\,s$^{-1}$, each at projected 
distances of 120 kpc away to the east and southeast
respectively.  The nearest H\,{\sc i}-rich galaxy within 200 
km\,s$^{-1}$ is NGC 4192A at a projected distance of 290 kpc \citep{Davies04}. 
If our detection were tidal debris, then it would have to have been drawn out 
on a timescale
of 16 kpc/200 km\,s$^{-1}$ (200 km\,s$^{-1}$ being the typical velocity width
within a single Arecibo beam), or $6 \times 10^{7}$ yr.  It follows that the 
interacting galaxies which generated it must still be close enough that they 
could have been near VIRGOHI21 $6 \times 10^{7}$ yr ago.  For the two 
apparently nearest galaxies (above), that would imply relative projected speeds
of greater than 1500 km\,s$^{-1}$.  This is very high compared to the 
velocities
of galaxies in the outskirts of Virgo (velocity dispersion $\sim 700$
km\,s$^{-1}$) where VIRGOHI21 appears to be 
situated \citep{Davies04}, and far too high to favor significant tidal 
interactions \citep[e.g.][]{Toomre72,Barnes92}.  Furthermore, the 
observations of VIRGOHI21
show higher velocities to the north and lower velocities to the south, while
the two nearby galaxies are placed with the one at a higher velocity to the
south and the one with a lower velocity slightly north of due east -- the 
opposite
sense to that expected if the velocity gradient in VIRGOHI21 were due to a
tidal interaction between them.

If a tidal origin is thus excluded what alternative explanations are 
compatible with the wide velocity width?  Several narrow-line higher 
column-density clouds at
different velocities lined up in the beam?  Clouds like this are often 
associated with tidal debris as the filamentary structure breaks up into 
separate H\,{\sc i} clouds which may form tidal dwarf galaxies 
\citep[e.g.][]{Hunsberger96}.  Such clouds should have
been detected by our VLA observations; that they were not implies that this
is an unlikely explanation.  Another possibility is that the gas is not bound 
-- but then it should have dispersed in the same short time-scale of $6 \times 
10^{7}$ yr.  Given the dynamical timescale of the cluster of $\sim 10^9$ 
yr this possibility seems unlikely.  The Galactic extinction at this point in 
the sky is only 0.15 mags in $B$ band
\citep{Schlegel98}, therefore it is very unlikely that a galaxy has been
hidden by obscuration. As the other possibilities seem so unlikely, one is
left with the hypothesis that the gas in VIRGOHI21 is gravitationally bound 
and moving in stable, bound orbits which prevent shocking -- rotation of 
course comes to mind, as in a flattened disk -- a model which is not 
inconsistent with the spectra.  If the system is bound, then its 
dynamical mass $M_{dyn} \equiv R_{HI} \times \Delta V^2 / G$ is greater than
$9\times 10^{10}$ solar masses (with $R_{HI} \geq$ 8 kpc and $\Delta V$ = 220 
km\,s$^{-1}$), not atypical of a rotating galaxy, though its $M_{dyn}/M_{HI} > 
400$ is about 5 times higher than normal spirals.

From the well known Tully-Fisher correlation between rotational velocity
and luminosity, calibrated in the Virgo Cluster by \citet{Fouque90}, our 
H\,{\sc i} detection, if indeed it is a bound system, should correspond to a 
galaxy with an absolute $B$ magnitude of -19.  At the distance of the Virgo 
Cluster this would correspond to a 12th magnitude galaxy, which would normally
be extremely prominent at optical wavelengths.  VIRGOHI21 appears to 
be a massive object not containing the expected bright galaxy.

It has been proposed that there 
is an H\,{\sc i} column-density threshold ($\sim 10^{20}$ cm$^{-2}$) 
below which star formation ceases to occur \citep{Toomre64,Martin01}. 
The mean column-density across our central beam at Arecibo is somewhat lower 
than this, at $4 \times 10^{19}$ cm$^{-2}$ and our VLA observations set an 
upper limit to the column-density of $10^{20}$ cm$^{-2}$.  This low 
column-density provides an explanation for
the lack of an optical counterpart:  this may be a dark 
galaxy that has failed to form stars because the low disk surface-density 
prevents fragmentation of the gas, i.e. it does not satisfy Toomre's 
criterion \citep{Verde02,Toomre64}.

If such dark objects exist in significant numbers, then why has it taken until 
now to detect one?  VIRGOHI21-like objects could only have been detected by
H\,{\sc i} surveys which meet three criteria: (1) that they are `blind', rather
than targeted at previously-identified objects (which, by definition, are not
`dark'); (2) that they have 5$\sigma$ 
column-density sensitivity to galaxies with velocity widths $\sim 200$ 
km\,s$^{-1}$ at the $5 \times 10^{19}$ cm$^{-2}$ level (which rules out
older H\,{\sc i} surveys); (3) that they have complete optical follow-up 
observations to deep isophotal limits.  While there have been many blind
surveys, the only ones to meet the second criterion are HIPASS \citep{Meyer04},
HIJASS \citep{Lang03}, AHISS \citep{Zwaan97}, HIDEEP \citep{Minchin03} and 
VIRGOHI \citep{Davies04} and of these, only the last three satisfy the third
criterion -- that they have complete optical follow-up data.
The HIPASS Bright Galaxy Catalogue \citep{Koribalski04} (BGC; peak flux $>$
116 mJy), which has recently been used to make the most accurate determination 
to date of the H\,{\sc i} mass function \citep{Zwaan03}, would not have
detected VIRGOHI21 unless it were within six Mpc -- a relatively small distance
and very close to the level where it would be impossible, in velocity-space, 
to distinguish a truly isolated H\,{\sc i} cloud from one associated with The
Galaxy.   Taking the volume in which VIRGOHI21 would have been detected
(in HIDEEP, VIRGOHI and AHISS) leads to a global density of $\sim 0.02$ 
Mpc$^{-3}$, equivalent to a contribution to the cosmic density of $\Omega\simeq
0.01$.  To have had more than one detection would therefore imply a very 
significant contribution to the cosmic density.

\section{Conclusions}

In the very nature of things it would be difficult to make an indisputable 
claim
to have found a dark galaxy, particularly when past claims to that effect have 
quickly been ruled out by subsequent observations (either of a dim underlying 
galaxy or of bridging connections to nearby visible companions).  Nevertheless,
VIRGOHI21 passes all of the careful tests we have been able to set for it, 
using the best equipment currently available.  Far longer VLA observations
might help -- but the very low column density and broad velocity width will 
make VIRGOHI21 an extremely challenging test for any current interferometer.  
And if every deep 21-cm detection without an optical counterpart is dismissed 
out of hand as debris -- without considering the timing argument we
present in Section 3 (which can be used to exclude all previous claims) -- one 
is in effect 
ruling out, by definition, the detection of any dark galaxy at 21-cm.  We 
cannot of course be certain, but VIRGOHI21 has turned up in one of the two 
extremely deep 21-cm surveys where you could most reasonably expect to find a 
dark galaxy, and meets all of the criteria we can, in practice, set for such an
elusive but potentially vital object today: in particular such a broad observed
velocity width cannot itself be attributed to a tidal interaction if the 
putative interacting galaxies are well outside the Arecibo beam -- as they 
surely must be here.  Future deep H\,{\sc i} surveys 
could reveal a population of such galaxies; with colleagues we are planning 
such at Arecibo, Jodrell Bank and Parkes.

\clearpage

\clearpage

\clearpage

\end{document}